# Quantum–Classical Iterative Refinement for the Pressure-Poisson Equation on NISQ Hardware

**Doyeol (David) Ahn**[1,2,*], **Byeongyong Park**[3], **Rho Shin Myong**[4], **Franz Klein**[5]

[1] Department of Electrical and Computer Engineering, University of Seoul, Seoul, Republic of Korea

[2] Singularity Quantum Inc., Newport Beach, CA, USA

[3] Center for Quantum Information Processing, University of Seoul, Seoul, Republic of Korea

[4] School of Mechanical and Aerospace Engineering, Gyeongsang National University, Jinju, Republic of Korea

[5] National Quantum Laboratory at Maryland (QLab), University of Maryland, College Park, MD, USA

[*] Corresponding author: dahn@uos.ac.kr

## Abstract

We present a hybrid quantum–classical iterative refinement for the pressure-Poisson equation of the incompressible Navier–Stokes projection method on noisy quantum hardware. Rather than solving the pressure equation once per time step, we solve it repeatedly for the normalized residual and scale each quantum direction by a least-squares fit to the physical residual. As a result, the error of the quantum solve enters as a contraction factor $\sin\theta_k$ of the residual rather than as an error floor. This removes the requirement of a state infidelity below $10^{-7}$ that was estimated for single-solve schemes. The noise of the hardware then sets the rate of convergence but not the attainable accuracy. It is this change of the role of the quantum solve, and not an improvement of the hardware, that allows the residual to converge. We use a boundary-consistent staggered (MAC) discretization, so that the pressure residual equals the velocity divergence left after projection up to the factor $\Delta t/\rho$.

On the IBM processor ibm_cleveland, a four-qubit (16 × 16) problem reached a relative residual below $10^{-7}$ within eight to ten corrections in all 20 runs of a pre-registered experiment ($8.6 \times 10^{-9}$ to $9.7 \times 10^{-8}$). A single solve with a comparable total number of shots stopped at 0.33–0.59 in every run. On the IonQ Forte processor, two corrections reduced the residual of the same problem by 89.2%, and on a manufactured-solution benchmark the relative residual reached the same tolerance of $10^{-7}$ in nine corrections ($5.6 \times 10^{-8}$), with a pressure error of $4.9 \times 10^{-8}$. In a six-qubit (64 × 64) pilot on ibm_cleveland, all three runs reached the same tolerance in 13 corrections, whereas single solves with a comparable number of shots stopped at 0.46–0.72. The per-correction contraction is of the same size as the infidelity of the reconstructed states. Finally, we carried out the variational training itself on ibm_cleveland, so that every sampled value of the cost function was obtained from hardware measurements and each correction started from zero parameters. The residual fell to $5.6 \times 10^{-4}$ in seven corrections, below the pre-specified tolerance of $10^{-3}$, at the rate predicted by a noise model and with 24 min of QPU time. To the best of our knowledge, these are the first converged pressure solutions obtained on quantum hardware.

# 1. Introduction

In the projection method for incompressible flow, the pressure is obtained at every time step from a Poisson equation whose source depends on the intermediate velocity [Chorin 1968]. The pressure is a nonlocal functional of the instantaneous velocity field, and the Poisson solve is usually the most expensive part of each step [Doering and Gibbon 1995]. The pressure acts as the Lagrange multiplier of the incompressibility constraint, so that the accuracy of the Poisson solve sets the accuracy of the constraint. If the pressure equation is solved only approximately, the velocity after projection keeps a divergence proportional to the residual. The pressure then does spurious work on the flow, and its effect on the kinetic energy accumulates over the time steps (Appendix A.4). A pressure solver is therefore useful only if its residual can be reduced to a prescribed tolerance. The Poisson solve is thus a natural, but demanding, target for quantum linear-system algorithms.

Quantum algorithms for partial differential equations have been developed actively, and several of them have appeared in this journal. Jin et al. [2023] analyzed the time complexity of quantum algorithms based on linear representations of nonlinear ordinary and partial differential equations. Jin et al. [2024] treated physical boundary and interface conditions with the Schrödingerisation approach. Schalkers and Möller [2024] constructed an efficient and fail-safe quantum algorithm for the transport equation. Mamun et al. [2027] formulated a unitary quantum lattice Boltzmann method for canonical flow and thermal transport. Meng et al. [2026] proposed an end-to-end quantum simulation of rapidly distorted turbulence, from the preparation of the initial state to the measurement of the turbulence statistics. In general, these algorithms target fault-tolerant devices, and their accuracy and cost are assessed by complexity estimates or by classical emulation of the circuits. As Meng et al. [2026] pointed out, the extraction of the solution from the quantum state is an essential part of the total cost.

Fault-tolerant algorithms such as HHL require circuit depths far beyond current hardware. The same holds for pure quantum algorithms for nonlinear flow equations; for example, the nonlinear Burgers equation for high-speed compressible flow has been solved with block encoding, the quantum Hadamard product and the linear combination of unitaries, so far on simulators only [Esmaeilifar et al. 2024]. Variational algorithms, such as the variational quantum eigensolver (VQE) and the variational quantum linear solver (VQLS), are shallow enough for noisy intermediate-scale quantum (NISQ) devices, but their output carries an error set by gate and readout noise [Bravo-Prieto et al. 2023; Liu et al. 2021]. Song et al. [2025] combined the Chorin projection with a VQE pressure solve, algebraic multigrid (AMG) preconditioning and an efficient sign reconstruction (HTree). Their state preparation and readout reached fidelities of 95–99% on IBM hardware. However, the maximal residual remained near $10^{-1}$, and increasing the number of shots from $10^5$ to $10^8$ did not help. They estimated that convergence requires a state infidelity $1 - F \leq 10^{-7}$ and concluded that current NISQ hardware is too noisy for explicit time stepping. Their maximal residual measures the change of the velocity between successive time steps, and it differs from the relative residual of the linear system used in the present work. Other variational Poisson solvers have been trained on hardware only for small systems or with low accuracy. Ali and Kabel [2023] trained a variational solver for a five-qubit, one-dimensional Poisson problem on IBM hardware and reached a state fidelity below 0.2. Tosti Balducci et al. [2024] trained the VQLS on IBM hardware for systems of up to 4 × 4. To the best of our knowledge, a converged pressure solution has not been obtained on quantum hardware before the present work.

The reason for this gap is structural. In a single-solve scheme, the measured quantum state is used as the solution itself. Its error is roughly the square root of the state infidelity. This error is dominated by a bias from gate and readout noise, and it does not decrease with the number of shots. The residual therefore stops at a floor set by the hardware. A converged pressure solution on present hardware thus requires a change of the role of the quantum solve rather than a better device.

In the present work, we ask a different question. Rather than asking how accurate a single quantum solve must be, we ask whether a solve of fixed, moderate accuracy can be used repeatedly so that the residual decreases geometrically. This is the idea of iterative refinement, in which a low-precision solver is combined with a residual computed in high precision [Wilkinson 1963; Carson and Higham 2018]. We show that the convergence barrier reported by Song et al. is a property of the single-solve structure and not of the hardware. The key element is a division of roles. The quantum processor supplies only a normalized direction, while the residual, the scale and the update are computed exactly on the classical side. The quantum state carries no information on the norm and the global sign of the solution, and the least-squares scale fitted to the physical residual recovers both. With this scale, the noise of the quantum solve no longer limits the attainable accuracy. It only sets the rate of convergence, through a contraction factor $\sin \theta_k$ per correction, and any direction with a nonzero overlap with the residual reduces the residual. The requirement on the hardware thus changes from a state infidelity of $10^{-7}$ to a contraction factor below one. On present hardware the contraction factors were well below one, and eight to 13 corrections were enough to reach a relative residual of $10^{-7}$. This is why a converged pressure solution can be obtained on present hardware. Compared with the simulator study of Saito et al. [2023], the present work carries the refinement to quantum hardware and to the pressure-Poisson equation of incompressible flow, and it relates each reduction of the residual to a reduction of the velocity divergence. We use the same real-amplitude ansatz, classical state-vector training and HTree readout, so that the difference lies in the solution structure and not in the quantum circuit. In addition, we show that the classical training can be replaced by training on the quantum processor itself.

It is shown that (1) with a least-squares scale fitted to the physical residual, each correction contracts the residual by $\sin\theta_k$, where $\theta_k$ is the angle between the residual and the image of the quantum direction; (2) with a boundary-consistent MAC discretization, the residual equals the velocity divergence after projection up to $\Delta t/\rho$; (3) on ibm_cleveland the residual reached below $10^{-7}$ in all 20 runs of a pre-registered four-qubit experiment, whereas a single solve with a comparable total number of shots stopped at 0.33–0.59, and the same held in a six-qubit pilot with 13 corrections per run; (4) on the IonQ Forte trapped-ion processor the same contraction behavior is observed, and a manufactured-solution benchmark reached the tolerance of $10^{-7}$ in nine corrections; and (5) with the variational training carried out on ibm_cleveland, so that every sampled cost value is obtained from hardware measurements, the residual fell below $10^{-3}$ in seven corrections from zero initial parameters.

The paper is organized as follows. Section 2 describes the method. Section 3 describes the hardware experiments, and Section 4 reports the results. Section 5 discusses the relation to single-solve schemes and the limitations, Section 6 concludes, Appendix A summarizes the governing equations and the derivations used in Section 2, and Appendix B gives the circuits executed on hardware.

# 2. Method

## 2.1 Projection step and pressure operator

We consider the incompressible Navier–Stokes equations for a fluid of constant density $\rho$ and kinematic viscosity $\nu$,

$$\nabla \cdot \mathbf{u} = 0, \qquad \frac{\partial \mathbf{u}}{\partial t} + (\mathbf{u} \cdot \nabla)\mathbf{u} = -\frac{1}{\rho}\nabla p + \nu\nabla^2\mathbf{u}. \tag{1}$$

In dimensionless form $\nu$ is replaced by 1/Re, where Re is the Reynolds number. In the projection method of Chorin [1968], one time step of length $\Delta t$ is split into a prediction without the pressure gradient, a pressure-Poisson solve and a velocity correction,

$$\frac{\mathbf{u}^* - \mathbf{u}^n}{\Delta t} = -(\mathbf{u}^n \cdot \nabla)\mathbf{u}^n + \nu \nabla^2 \mathbf{u}^n, \tag{2}$$

$$\nabla^2 p^{n+1} = \frac{\rho}{\Delta t} \nabla \cdot \mathbf{u}^*, \tag{3}$$

$$\mathbf{u}^{n+1} = \mathbf{u}^* - \frac{\Delta t}{\rho} \nabla p^{n+1}. \tag{4}$$

The Poisson equation enforces $\nabla \cdot \mathbf{u}^{n+1} = 0$. It is the only step that we pass to the quantum processor; the prediction and the correction are computed classically. The derivation of these steps, the discretization and the identities used below are given in Appendix A.

We discretize these steps on a staggered (MAC) grid of $N = n_x n_y$ pressure cells (Appendix A.2). The pressure then satisfies the discrete Poisson equation

$$Lp = b, \qquad L = -DG, \qquad b = -\frac{\rho}{\Delta t} Du^*, \tag{5}$$

where $D$ and $G$ denote the discrete divergence and gradient, and $L$ is the Neumann Laplacian with consistent wall boundary conditions. The pressure is determined only up to an additive constant [Doering and Gibbon 1995, Eq. (1.2.21)]. We fix the constant by the boundary-consistent operator $A = L + e_0 e_0^{\mathsf{T}} / \Delta x^2$. The rank-one term changes the operator, but for a compatible right-hand side the solution of Ap = b satisfies $p_0 = 0$ and Lp = b (Appendix A.3). After the projection $u = u^* - (\Delta t/\rho) Gp$, it can be shown (Appendix A.3) that

$$Du = -\frac{\Delta t}{\rho} r, \qquad r = b - Lp. \tag{6}$$

The pressure residual is therefore the velocity divergence left after projection, up to the factor $\Delta t/\rho$. Reducing the residual is identical to enforcing incompressibility.

## 2.2 Variational solve and readout

For a normalized right-hand side $|\hat{b}\rangle$, we prepare $|\psi(\theta)\rangle = U(\theta)|0\rangle$ with a real-amplitude ansatz (RealAmplitudes, five repetitions; 24 parameters and 15 CNOT gates for four qubits; six repetitions for the six-qubit pilot of Section 3.4). The parameters minimize the effective Hamiltonian $H = A^{\mathsf{T}}\big(I - |\hat{b}\rangle\langle\hat{b}|\big)A$ of Liu et al. [2021], with left AMG preconditioning as in Song et al. [2025]. The normalized energy of a trial state $\psi$ equals $\sin^2\varphi$, where $\varphi$ is the angle between $A\psi$ and $b$ (Appendix A.5). Unless stated otherwise, training uses an exact classical state-vector simulation with adjoint gradients, and the quantum processor prepares the trained state and reads it out. In Sections 2.6, 3.5 and 4.6 the training is carried out on the quantum processor. For problem P2 the direction is fitted with a different objective: with the Jacobi preconditioner J = diag(L), the parameters minimize the physical residual after the optimal scaling of the mean-free direction $J^{-1}\psi$, with two random starts of at most 300 L-BFGS-B iterations.

The amplitudes are real, so that the state is determined by its magnitudes and relative signs. We measure $n+1$ circuits: one in the computational basis, which gives $|\psi_x| = \sqrt{P_Z(x)}$, and one with a Hadamard gate on each qubit $j$, which gives $P_{H,j}(x_0) - P_{H,j}(x_1) = 2\psi(x_0)\psi(x_1)$ for basis states $x_0$, $x_1$ that differ only in bit $j$. The signs are propagated along a maximum spanning tree of these pairs (HTree). Readout errors are corrected with an empirical assignment matrix, followed by a projection onto the probability simplex.

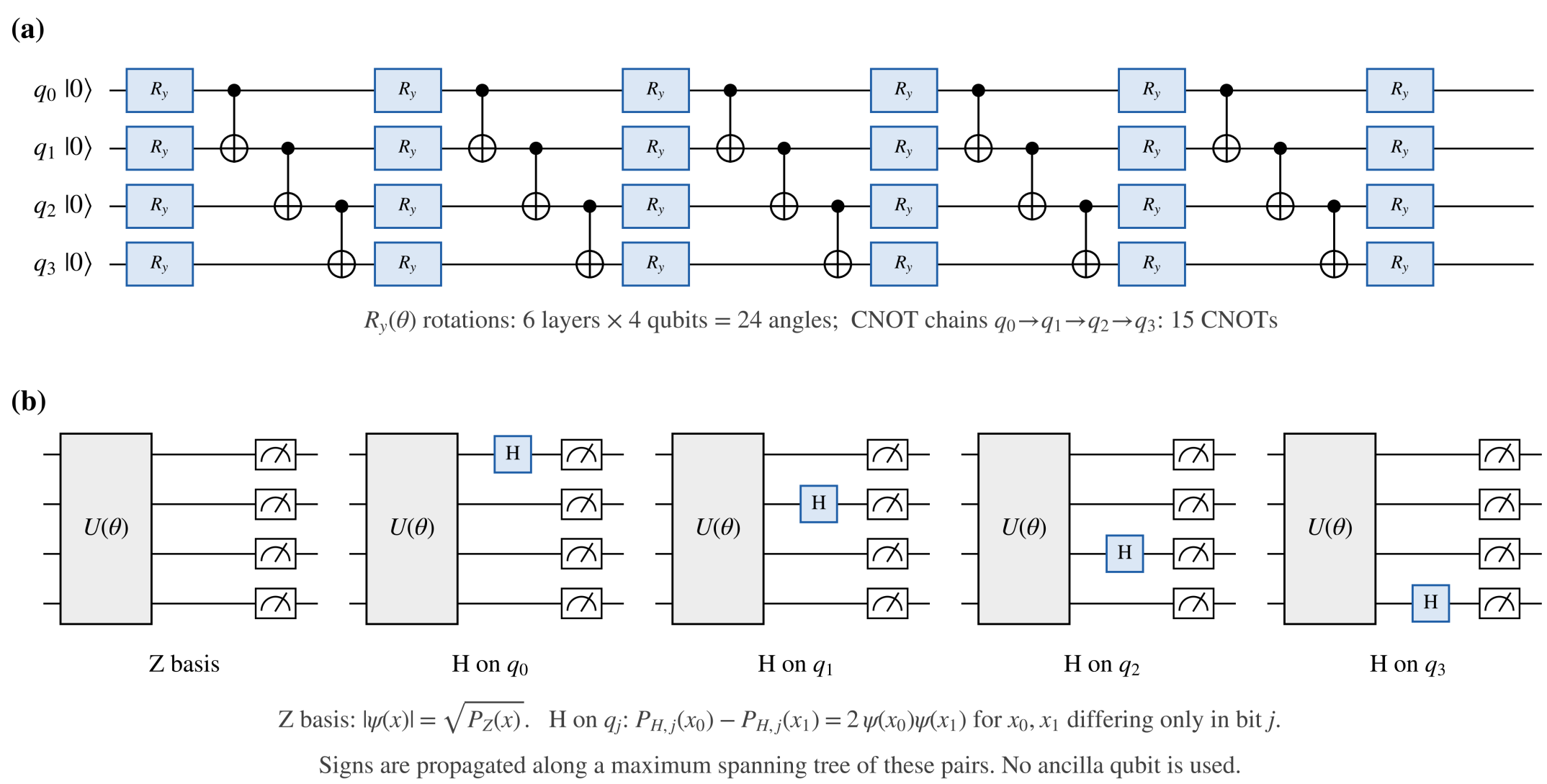

*Fig. 1. (a) Real-amplitude ansatz $U(\theta)$ with 24 trained angles and 15 CNOT gates. (b) The five HTree readout circuits measured at every correction.*

The logical circuit is the same on both processors. On ibm_cleveland the 15 CNOTs compile to 15 CZ gates on the linear chain of qubits 21–24, without SWAP gates. On IonQ Forte they compile to 15 ZZ gates, with native GPI and GPI2 single-qubit gates. The circuits executed on hardware are given in Appendix B.

## 2.3 Residual refinement

Let $p_k$ be the current pressure and $r_k = b - Lp_k$ its residual, with $p_0 = 0$. At each correction we train and measure the state for the normalized residual $r_k/\| r_k \|$ and obtain a direction $d$. We remove the amplitude of the pinned cell, $d^0 = d$ with $d^0[0] = 0$, and choose the scale that minimizes the physical residual,

$$\alpha_k = \frac{(Ld^0)\cdot r_k}{\| Ld^0 \|^2}, \qquad p_{k+1} = p_k + \alpha_k d^0, \qquad r_{k+1} = r_k - \alpha_k L d^0. \tag{7}$$

For P2, the reconstructed vector is first mapped by $J^{-1}$ and its mean is removed, and the pressure is shifted afterward to restore $p_0 = 0$. The global sign of the reconstructed state cancels in $\alpha_k d^0$. The refinement stops when $\| r_k \|/\| b \| \leq \tau$ or when a cap of $K$ corrections is reached.

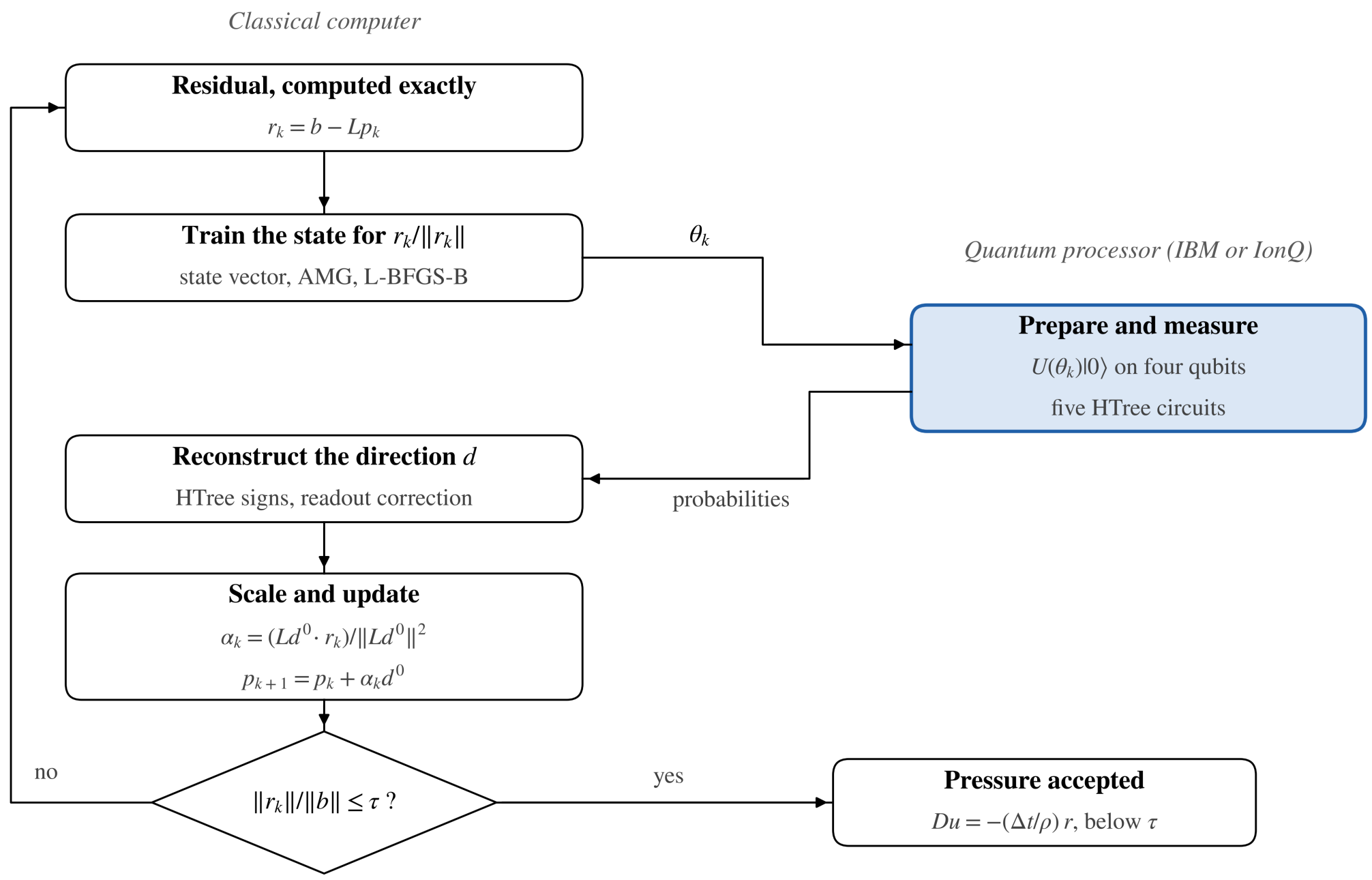

*Fig. 2. The hybrid quantum–classical refinement loop. The residual, the scale and the update are computed exactly on the classical side; the quantum processor supplies one normalized direction per correction.*

## 2.4 Contraction per correction

By substituting the optimal $\alpha_k$ into the residual update, we obtain

$$\| r_{k+1} \| = \| r_k \| \sin\theta_k, \qquad \cos\theta_k = \frac{|(Ld^0) \cdot r_k|}{\| Ld^0 \| \| r_k \|}, \tag{8}$$

where $\theta_k$ is the angle between the residual and the image of the quantum direction. The residual therefore never increases, and it decreases whenever $\theta_k < \pi/2$. After $K$ corrections, $\| r_K \| / \| b \| = \prod_k \sin\theta_k$.

This result can also be deduced from physical intuition only. Moving along a direction that makes an angle $\theta$ with the residual, by the best distance, leaves only the component perpendicular to that direction. Its magnitude is the original residual multiplied by $\sin\theta$. Repeating the step multiplies these factors. The contraction relation follows.

The requirement on the quantum solve thereby changes. In a single-solve scheme the solution error is fixed by the state infidelity, roughly $\sqrt{1-F}$ in direction, and the residual cannot fall below it. In the refinement scheme the same error appears as $\sin\theta_k$. A direction with $\sin\theta \approx 0.3$ reduces the residual by a factor of about $6 \times 10^{-6}$ in ten corrections.

## 2.5 Energy-based tolerance

In time stepping, the residual tolerance can be chosen from a physical criterion. The pressure work on the divergence left by an inexact solve acts as a spurious power $P_s = \int p \, \nabla \cdot \mathbf{u} \, d^d x$ in the kinetic-energy balance [Doering and Gibbon 1995, Eq. (1.4.24); Appendix A.4]. With $Du = -(\Delta t/\rho)\, r$ and the Cauchy–Schwarz inequality, $|P_s| \leq (\Delta t/\rho) \parallel p \parallel \parallel r \parallel$. In order to keep $P_s$ below a fraction $\varepsilon_E$ of the viscous dissipation rate $\rho\nu \parallel \nabla\mathbf{u} \parallel^2$, it is sufficient that

$$\frac{\parallel r^n \parallel}{\parallel b^n \parallel} \leq \tau_n = \frac{\varepsilon_E\, \rho^2 \nu \parallel \nabla\mathbf{u} \parallel^2}{\Delta t \ \parallel p \parallel \parallel b^n \parallel}. \tag{9}$$

The quantities on the right-hand side can be estimated classically from the previous step. In the present paper we report single-step results with a fixed $\tau = 10^{-7}$; the adaptive tolerance is used in the time-stepping study that follows.

## 2.6 On-device training

In Sections 3.5 and 4.6 the variational training is carried out on the quantum processor. Every sampled value of the cost function is then obtained from hardware measurements. For a parameter vector $\theta$, we measure the five HTree circuits, correct the readout errors and reconstruct the real state $\hat{\psi}(\theta)$ as in Section 2.2. The cost is the quadratic form

$$E(\theta) = \hat{\psi}^{\mathsf{T}} \widetilde{H}\, \hat{\psi}, \qquad \widetilde{H} = \tilde{A}^{\mathsf{T}}\big(I - \tilde{b}\,\tilde{b}^{\mathsf{T}}\big)\tilde{A}, \tag{10}$$

where $\tilde{A} = M^{-1}A$, $\tilde{b} = M^{-1}r_k / \parallel M^{-1}r_k \parallel$ and $M^{-1}$ is the left preconditioner of Section 2.2. The classical computer only reconstructs the state and evaluates this quadratic form. It uses neither the exact solution nor the ideal state of the circuit.

We minimize $E$ with the Rotosolve method [Nakanishi et al. 2020; Ostaszewski et al. 2021]. Each angle $\theta_j$ enters the ansatz through a single $R_y$ gate. For fixed values of the other angles, the ideal expectation value is therefore a sinusoid of period $2\pi$ in $\theta_j$. We use this form as a model for the reconstructed cost, which also contains sampling noise and the nonlinear reconstruction. Three values determine it: the current cost $E_0$ and the costs $E_\pm$ measured at $\theta_j \pm \pi/2$. The minimum of the sinusoid gives the update

$$\theta_j \leftarrow \theta_j - \frac{\pi}{2} - \operatorname{atan2}(2E_0 - E_+ - E_-,\ E_+ - E_-), \tag{11}$$

$$E_0 \leftarrow \frac{E_+ + E_-}{2} - \sqrt{\left(E_0 - \frac{E_+ + E_-}{2}\right)^2 + \left(\frac{E_+ - E_-}{2}\right)^2}. \tag{12}$$

The angles are updated in the order $j = 0,1, \ldots, 23$, and the order is repeated for a second sweep. The cost at the initial parameters is measured once; after that, $E_0$ is the fitted minimum of the previous update. Each update therefore needs one job of ten circuits, the five HTree circuits at $\theta_j + \pi/2$ and at $\theta_j - \pi/2$. The training of one correction consists of 48 such jobs, and the trained state is then measured once more at a larger number of shots to give the direction $d$ of Section 2.3. No randomness enters the hardware path: every correction starts from $\theta = 0$, and the updates are deterministic for given counts.

The training starts anew at each correction and does not use the parameters of the classical training. The refinement of Section 2.3 is unchanged.

# 3. Hardware experiments

## 3.1 Problems

Problem P1 is a 4 × 4 pressure-Poisson equation (four qubits, $N = 16$, $\Delta x = 0.25$) with a fixed right-hand side $b$, $\| b \| = 2.533$. The same $b$ is used on both processors, so that the IBM and IonQ results can be compared directly. The classical solution is used only for diagnostics after each run.

Problem P2 is a manufactured-solution MAC benchmark with Neumann boundary conditions, four qubits and a Jacobi preconditioner. Its exact pressure is known, so that the pressure error can be reported in addition to the residual. P2 was run on IonQ only.

## 3.2 IBM ibm_cleveland

The pre-registered experiment was run on 14 October 2026 (UTC) on physical qubits 21–24 of ibm_cleveland (IBM Quantum System One, Cleveland Clinic), with the Qiskit Runtime SamplerV2 in job mode. The circuits were transpiled at optimization level 1 with a fixed layout. Each refinement correction used the five HTree circuits at 4,000 shots each (20,000 shots per correction). Readout errors were corrected with a full 16 × 16 assignment matrix measured once per arm (16 circuits × 4,000 shots).

The protocol, the seeds and the stopping rules were fixed before submission. We ran 20 pairs (seeds 91000–91019) in four blocks of five. In each pair, the refinement arm ran with $\tau = 10^{-7}$, a cap of $K = 16$ corrections and a stop rule for three successive contractions above 0.7. The control arm (direct, fixed) used a single solve with the five circuits at 34,000 shots each. Its total of 234,000 shots, including the calibration, lies within the range of the refinement arms (224,000–264,000). This control isolates the effect of the solution structure from the effect of the shot budget. The run stopped automatically after pair 11, the first pair of the third block, because the recorded calibration of the qubit chain had changed. It was resumed for pairs 12–20 from the saved records. At the resumption, the stop rule was restricted to changes of the calibration values, so that a new calibration date alone did not stop the run. The calibration values of the chain were unchanged within each block, and no pair was repeated or excluded. An earlier pilot on 8 October 2026, with three pairs and a control matched shot for shot, showed the same behavior (Table 1).

## 3.3 IonQ Forte

The IonQ experiments used the Forte-1 trapped-ion processor through the IonQ v0.4 API, accessed through the National Quantum Laboratory at Maryland (QLab). The trained circuits were compiled to the native gate set, and each HTree circuit was run at 1,000 shots with debiasing off. The API returns probabilities rather than counts. All 16 keys were present, every probability was a multiple of 1/1,000, and no renormalization or imputation was applied. The bit order ($q_0$ first, width set by the highest qubit used) was verified separately by single-X diagnostic circuits. No assignment-matrix calibration was run on IonQ.

For P1 we ran two adaptive corrections; the second state was trained on the measured residual of the first. For P2 we ran the refinement to the tolerance $\tau = 10^{-7}$, which was reached in nine corrections, each trained on the residual reconstructed from the measured probabilities of the previous one. No matched-shot control was run on IonQ.

## 3.4 IBM six-qubit pilot

Problem P3 is the 8 × 8 version of the projection step (six qubits, $N = 64$, $\Delta x = 1/8$). Its right-hand side is the predictor of the lid-driven cavity at Re = 100 and $t = 4.0$ with $\Delta t = 0.1$, with $\| b \| = 24.82$. A noise-free training study showed that five repetitions of the ansatz stop in local minima at a state fidelity of about 0.91, and that more iterations do not help. We therefore used six repetitions (42 parameters, 30 CNOT gates) and kept the best of five random starts with 1,000 L-BFGS-B iterations each.

The protocol was fixed before submission from a noise model of qubits 21–26. The model scaled the reported gate errors by a single factor of 4.5. With this factor it reproduced the 185 four-qubit hardware contractions (simulated first and later medians 0.406 and 0.141, against 0.405 and 0.140). We ran three pairs (seeds 92000–92002) on 15 October 2026 (UTC) on physical qubits 21–26, where each circuit compiled to 30 CZ gates without SWAP gates. The refinement arm used the seven HTree circuits at 4,000 shots each, a full 64 × 64 assignment matrix (64 circuits × 4,000 shots), $\tau = 10^{-7}$, a cap of $K = 24$ and a stop rule for three successive contractions above 0.9. The threshold was raised from 0.7 because the model predicted first contractions near 0.7. The control arm used a single solve with the seven circuits at 62,000 shots each, 690,000 shots in total including the calibration. In every pair the refinement arm ran before the control arm, rather than in the randomized order of the protocol.

## 3.5 IBM on-device training

The on-device experiment used problem P1 on ibm_cleveland. The protocol, the predictions and the stopping rules were fixed before any on-device measurement. The run was carried out on 15 and 16 October 2026 (UTC) on the linear chain of physical qubits 9–12, which was selected at run time by the same calibration score as in Section 3.2. The Qiskit Runtime SamplerV2 was used in job mode, without a session. Each submitted circuit contained 15 CZ, 78 $R_z$ and 58 SX gates, with a depth of 56, and no SWAP gates.

A full 16 × 16 assignment matrix (16 circuits × 4,000 shots) was measured once at the start. The training used 500 shots per circuit, so that one correction used 242,500 training shots in 49 jobs. The trained state was measured with the five HTree circuits at 4,000 shots each. One correction therefore consisted of 50 jobs and 262,500 shots.

The tolerance was set to $\tau = 10^{-3}$, with a cap of $K = 10$ corrections and a stop rule for three successive contractions above 0.9. A larger tolerance than in Sections 3.2–3.4 was chosen because one correction costs 50 jobs instead of one. Before each correction, the run continued only if the QPU time reported by IBM plus a reserve of 230 s stayed within a budget of 6,000 s.

The noise model of Section 3.4, applied to this protocol, predicted that 20 of 20 simulated runs reach $10^{-3}$ in 5–9 corrections (median 7). It predicted a first contraction of 0.43, a median later contraction of 0.35 and a median QPU time of 1,442 s.

The run was stopped twice by an automatic rule that halted the submission when the calibration values of the chain changed. It was resumed from the saved jobs without resubmission. At the second resumption, the stop on a calibration change was replaced by a record of the change. This is the only deviation from the protocol. The readout calibration measured at the start was used throughout the run.

# 4. Results

## 4.1 IBM: refinement against the matched single solve

All 20 refinement runs reached the tolerance of $10^{-7}$, in eight to ten corrections. The single solves with a comparable total number of shots stopped at relative residuals of 0.33–0.59 (Fig. 3, Table 1).

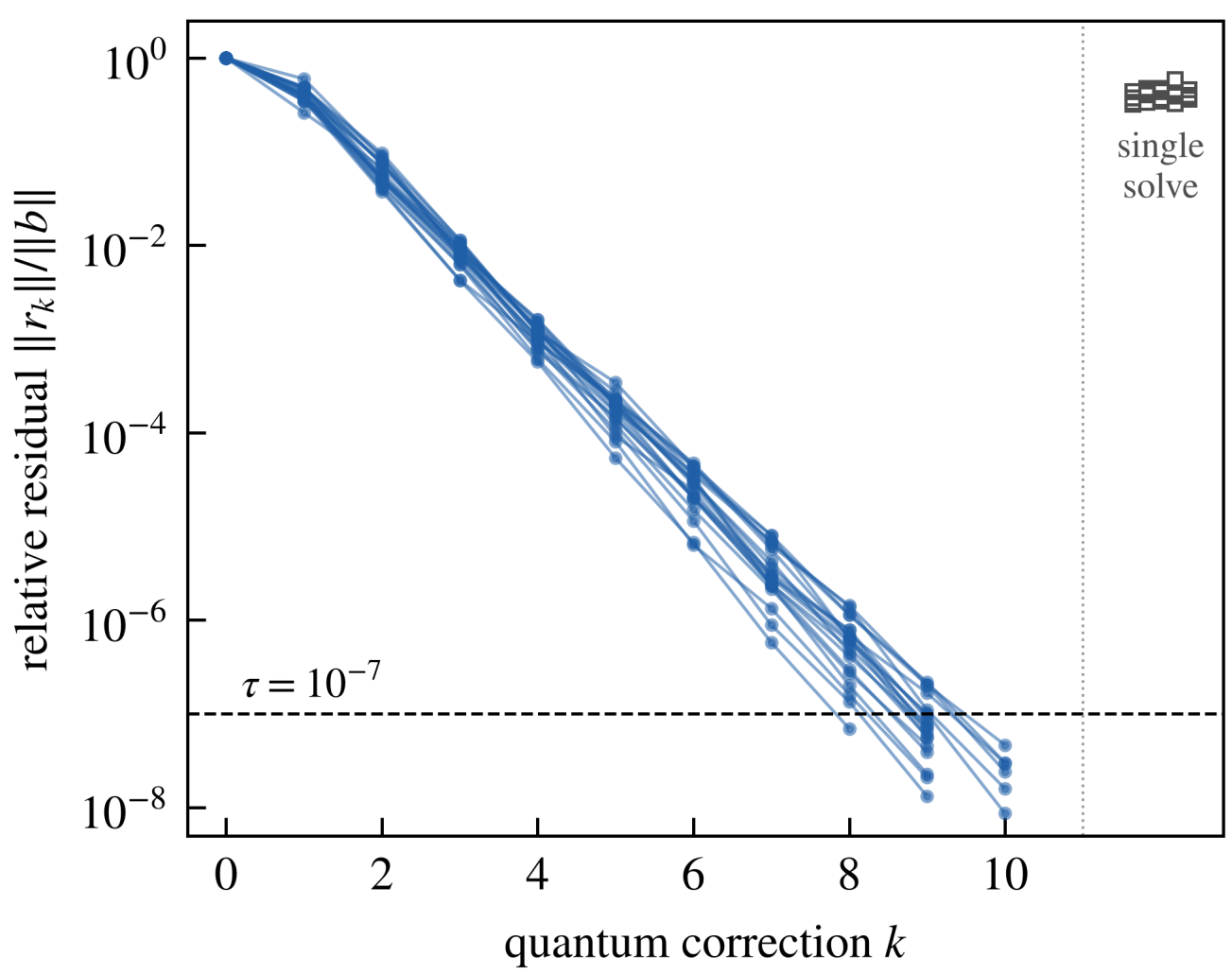

*Fig. 3. Relative residual of problem P1 against the number of quantum corrections (IBM ibm_cleveland, pre-registered 20-pair experiment of 14 October 2026). Squares at the right: the single solves with a comparable total number of shots.*

Table 1. IBM ibm_cleveland, problem P1. Total shots include the 64,000-shot readout calibration of each arm.

| Experiment | Runs reaching $10^{-7}$ | Corrections | Final relative residual, median (range) | Total shots per refinement arm | Single solve, median (range) |
|---|---|---|---|---|---|
| Pre-registered, 14 Oct 2026 | 20/20 | 8–10 (median 9) | $4.6 \times 10^{-8}$ ($8.6 \times 10^{-9}$ – $9.7 \times 10^{-8}$) | 224,000–264,000 | 0.40 (0.33–0.59) at 234,000 shots |
| Pilot, 8 Oct 2026 | 3/3 | 8–9 | $3.9 \times 10^{-8}$ ($3.4 \times 10^{-8}$ – $7.6 \times 10^{-8}$) | 224,000–244,000 | 0.35 (0.31–0.37) at the same shots |

In the pre-registered experiment, the first correction contracted the residual by 0.26–0.60 (median 0.40). The 165 later corrections had a median contraction of 0.140 (range 0.070–0.290), and no correction approached the stop threshold of 0.7. In the pilot, the overlaps of the reconstructed states with the ideal trained states in one run were 0.966–0.984, so that $\sqrt{1-F} = 0.13$–$0.18$; this is the size of the observed $\sin\theta_k$. We recomputed all 40 arms of the pre-registered experiment independently from the raw counts. The recomputed residuals agreed with the recorded ones, and the final pressures of the refinement arms agreed to $10^{-16}$.

The single solve and the refinement used the same circuits, the same processor and a comparable shot budget. One stopped and the other converged. The difference therefore comes from the solution structure and not from the shots or the hardware.

## 4.2 IonQ Forte

On problem P1, two adaptive corrections contracted the residual by 0.347 and 0.313, so that the relative residual fell to 0.108 (89.2% reduction). The first contraction lies within the IBM range (0.26–0.60). The second is larger than the IBM median of later corrections (0.140). The IonQ runs used 1,000 shots per circuit and no assignment-matrix calibration, against 4,000 shots and a full calibration on IBM. The same refinement thus works on superconducting and trapped-ion hardware without modification.

On problem P2, the refinement reached the tolerance of $10^{-7}$ after nine corrections, with a relative residual of $5.61 \times 10^{-8}$ (Fig. 4, Table 2). The relative error of the pressure against the manufactured solution fell from $7.47 \times 10^{-2}$ after the first correction to $4.88 \times 10^{-8}$ after the ninth, by nearly the same factors as the residual. The contraction varied between 0.11 and 0.28 without a systematic trend with the size of the residual; its geometric mean over nine corrections, 0.156, is close to the IBM median of 0.140. We attribute the spread to shot noise at 1,000 shots per circuit and to the absence of readout calibration.

Table 2. IonQ Forte-1, 1,000 shots per circuit, five circuits per correction.

| Problem | Correction | Contraction | Relative residual | Pressure relative error |
|---|---|---|---|---|
| P1 | 1 | 0.347 | 0.347 | — |
| P1 | 2 | 0.313 | 0.108 | — |
| P2 | 1 | 0.137 | 0.137 | $7.47 \times 10^{-2}$ |
| P2 | 2 | 0.190 | $2.61 \times 10^{-2}$ | $1.45 \times 10^{-2}$ |
| P2 | 3 | 0.179 | $4.66 \times 10^{-3}$ | $2.37 \times 10^{-3}$ |
| P2 | 4 | 0.125 | $5.85 \times 10^{-4}$ | $4.98 \times 10^{-4}$ |
| P2 | 5 | 0.150 | $8.74 \times 10^{-5}$ | $4.41 \times 10^{-5}$ |
| P2 | 6 | 0.113 | $9.88 \times 10^{-6}$ | $7.92 \times 10^{-6}$ |
| P2 | 7 | 0.277 | $2.73 \times 10^{-6}$ | $1.65 \times 10^{-6}$ |
| P2 | 8 | 0.181 | $4.96 \times 10^{-7}$ | $3.29 \times 10^{-7}$ |
| P2 | 9 | 0.113 | $5.61 \times 10^{-8}$ | $4.88 \times 10^{-8}$ |

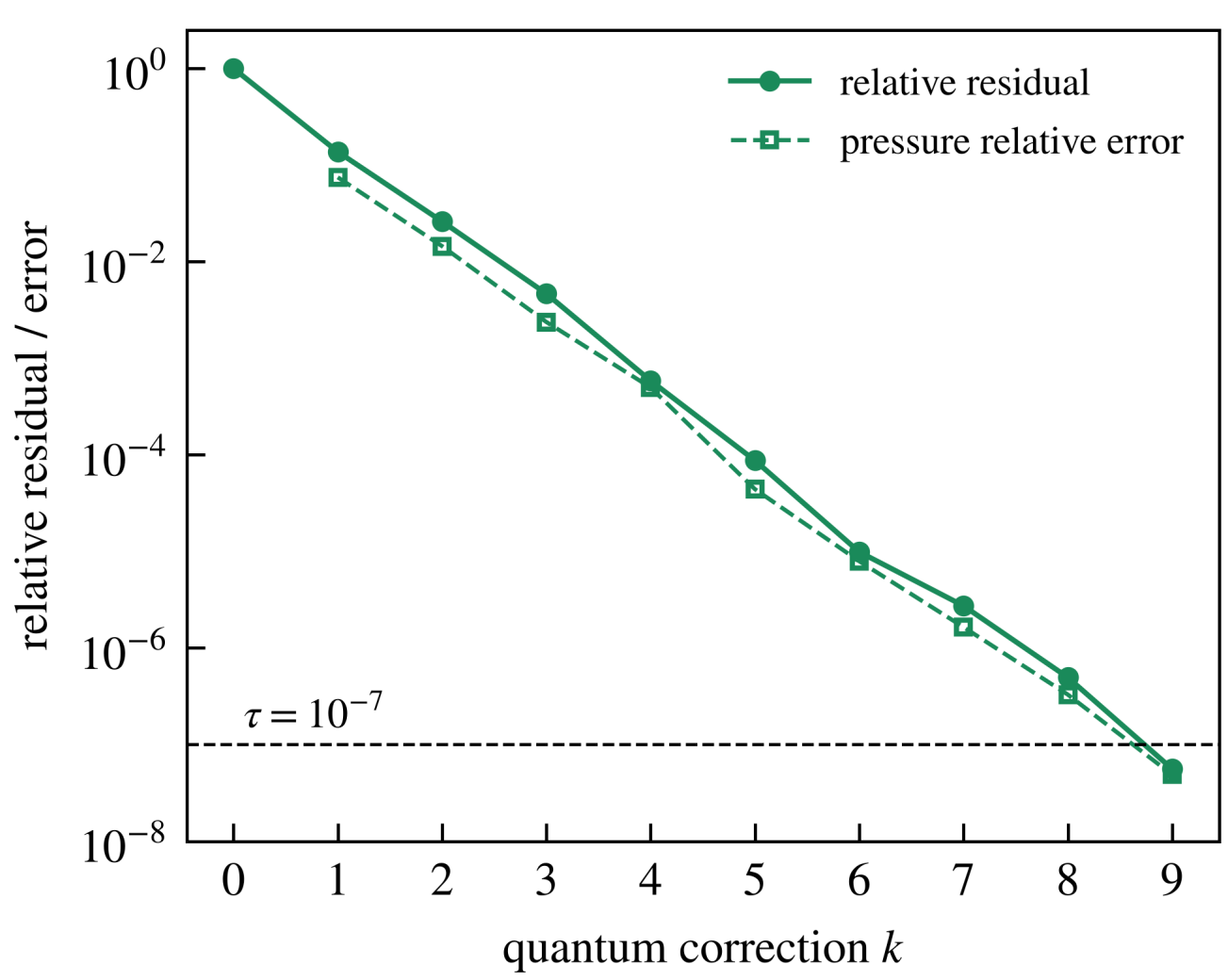

*Fig. 4. Relative residual and pressure error of problem P2 on IonQ Forte against the number of quantum corrections.*

## 4.3 Residual and divergence

The identity $Du = -(\Delta t/\rho)\, r$ held to $7 \times 10^{-17}$ on P1 and $2.3 \times 10^{-15}$ on P2 at every correction. On P1 the divergence L2 norm fell from 0.507 to 0.176 and 0.055 over the two IonQ corrections, by the same factors as the residual. Each correction therefore improves incompressibility by exactly its contraction factor.

## 4.4 Cost

The pre-registered IBM experiment, both arms of the 20 pairs, used 9.66 million shots and 3,180 s (53 min) of charged QPU time over 245 jobs; the count includes the jobs of a block that was interrupted and resumed. On IonQ, P1 used 10,000 shots and P2 used 45,000 shots over nine corrections. Both IonQ problems used 15 two-qubit ZZ gates per circuit, but the native compilation of P1 contained 336 single-qubit gates against 58 for P2. The on-device run of Section 4.6 used 1,440 s of charged QPU time over 351 jobs and 1.90 million shots. The complete pre-registered experiment of 40 arms thus required less than one hour of QPU time.

## 4.5 IBM six-qubit pilot

All three refinement runs of problem P3 reached the tolerance of $10^{-7}$ in 13 corrections, with final relative residuals of $5.2 \times 10^{-8}$, $8.0 \times 10^{-8}$ and $7.6 \times 10^{-8}$ (Fig. 5, Table 3). The relative error of the pressure against the exact solution of the discrete problem was 1.4–$2.3 \times 10^{-8}$. The single solves stopped at relative residuals of 0.72, 0.46 and 0.59.

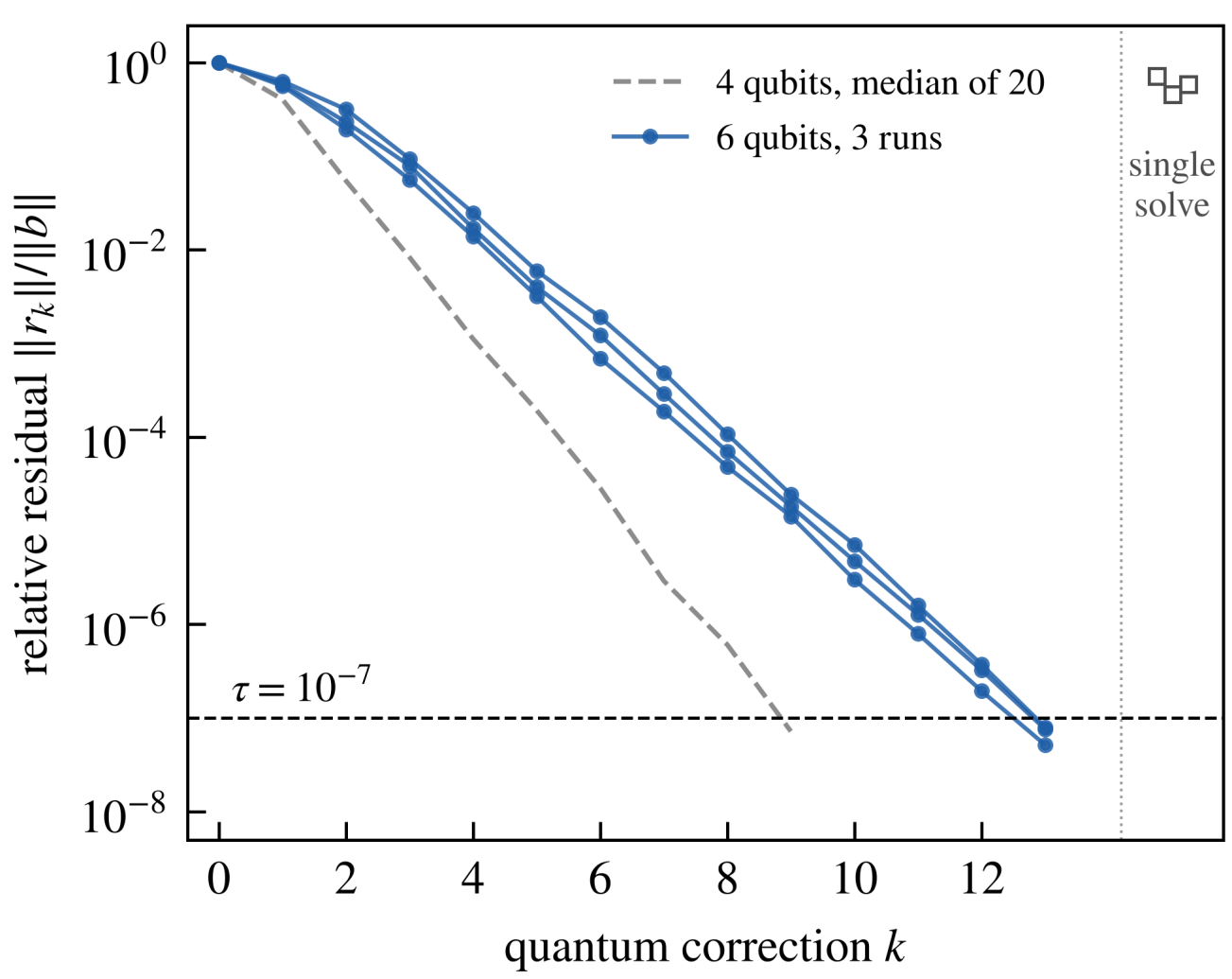

*Fig. 5. Relative residual of problem P3 (six qubits, 8 × 8) on ibm_cleveland against the number of quantum corrections. The dashed line is the median of the 20 four-qubit runs of Fig. 3; the squares at the right are the single solves at 690,000 shots.*

Table 3. IBM ibm_cleveland, problem P3, qubits 21–26. Refinement arms used 620,000 shots each, including the 256,000-shot readout calibration.

| Pair | Corrections | Final relative residual | Pressure relative error | First contraction | Later contraction, median | Single solve (690,000 shots) |
|---|---|---|---|---|---|---|
| 0 | 13 | $5.2 \times 10^{-8}$ | $1.4 \times 10^{-8}$ | 0.565 | 0.259 | 0.72 |

| Pair | Corrections | Final relative residual | Pressure relative error | First contraction | Later contraction, median | Single solve (690,000 shots) |
|---|---|---|---|---|---|---|
| 1 | 13 | $8.0 \times 10^{-8}$ | $2.3 \times 10^{-8}$ | 0.634 | 0.247 | 0.46 |
| 2 | 13 | $7.6 \times 10^{-8}$ | $2.2 \times 10^{-8}$ | 0.591 | 0.257 | 0.59 |

The 36 later corrections had a median contraction of 0.255 (range 0.210–0.500), and the geometric mean per run was 0.26–0.27. The overlap of the reconstructed states with the ideal trained states was 0.94–0.98 (median 0.965), so that $\sqrt{1-F} \approx 0.19$; this is again the size of the observed $\sin\theta_k$. Compared with four qubits, the later contraction increased from 0.140 to 0.255 and the number of corrections from about nine to 13, for four times as many unknowns. The noise model had predicted 15–17 corrections and a later contraction of 0.31, so that the hardware converged faster than the model predicted.

We recomputed all six arms independently from the raw counts. The contractions and the final residuals agreed with the recorded values, and the residual used to train each correction agreed with the residual reconstructed from the counts of the previous correction. The pilot used 3.93 million shots and 1,161 s of charged QPU time over 48 jobs. The pilot shows that the refinement carries over to six qubits, with four times as many unknowns, without any change of the method.

## 4.6 IBM on-device training

The refinement with on-device training reached the tolerance of $10^{-3}$ in seven corrections, with a final relative residual of $5.65 \times 10^{-4}$ (Fig. 6, Table 4). The first correction, which is a single on-device VQE solve, reduced the residual to 0.531. The six later corrections contracted the residual by 0.18–0.47, with a geometric mean of 0.32 and a median of 0.36. These values agree with the predictions of the noise model, 0.43 for the first contraction and 0.35 for the median later contraction. The number of corrections, seven, equals the predicted median.

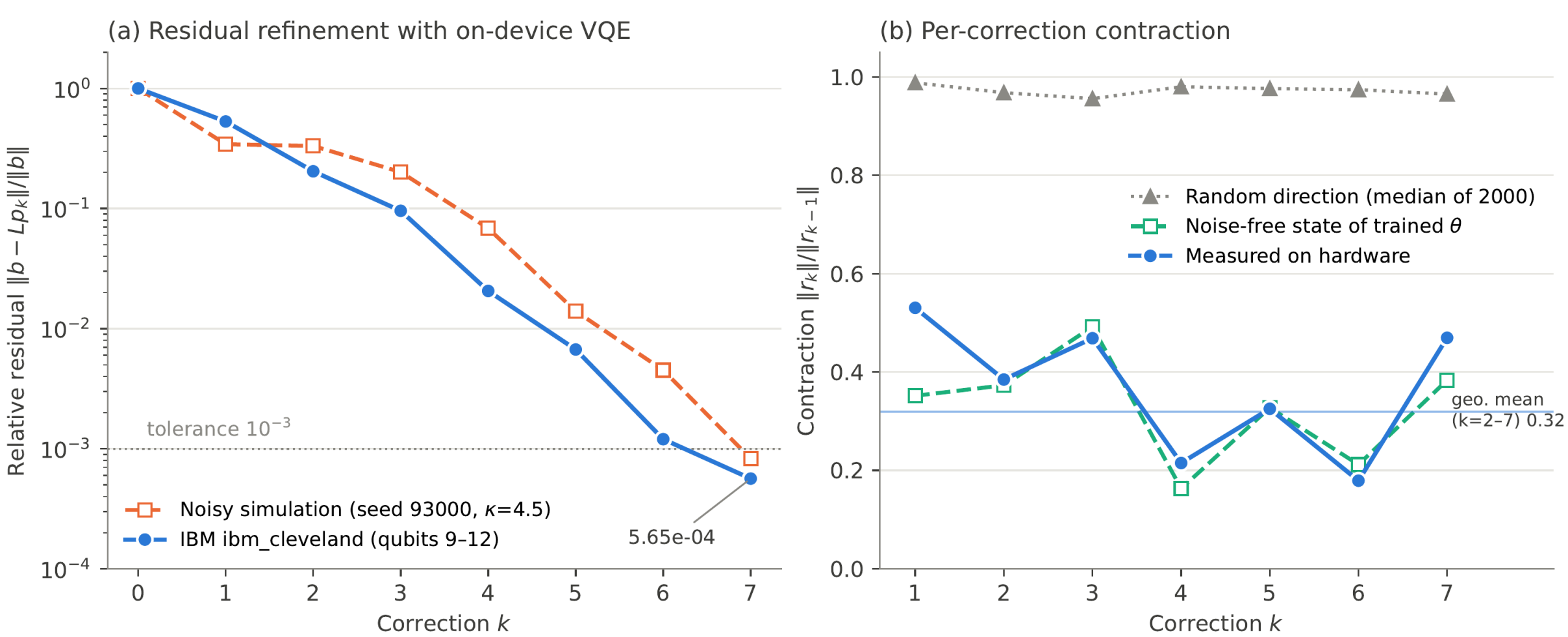

*Fig. 6. Refinement with on-device training on ibm_cleveland (problem P1, qubits 9–12). (a) Relative residual against the number of corrections, together with the noisy simulation of the pre-registered noise model. (b) Contraction per correction measured on hardware, the contraction of the noise-free state of the trained parameters, and the median contraction of 2,000 random directions. Every sampled cost value of the training was obtained from hardware measurements, and each correction started from θ = 0.*

The training reduced the measured cost from 0.99 at $\theta = 0$ to 0.042–0.215 at the trained parameters, measured with 4,000 shots (Table 4). After the run, we computed the fidelity of each trained state against the exact normalized solution. It was 0.77–0.97, which is lower than the fidelity of 0.966–0.984 obtained with classical training (Section 4.1). Even so, every correction contracted the residual, and none approached the stop threshold of 0.9.

In order to separate the training from the readout, we also computed the contraction that the noise-free state of the trained parameters would give (Fig. 6(b)). From the second correction on, the measured and the noise-free contractions differ by 0.003–0.087, and their geometric means are 0.32 and 0.30. A random direction gives a contraction of 0.99 (median of 2,000 directions). The contraction is therefore set mainly by the accuracy of the on-device training, and the readout at 4,000 shots adds little to it. The exception is the first correction, where the measured contraction was 0.53 against 0.35 for the noise-free state.

Table 4. IBM ibm_cleveland, problem P1, qubits 9–12, with the variational training on hardware. Each correction started from $\theta = 0$. The noise-free contraction and the fidelity are computed after the run from the trained parameters.

| Correction | Relative residual | Contraction, hardware | Contraction, noise-free state | Measured cost at trained $\theta$ | Fidelity of trained state |
|---|---|---|---|---|---|
| 1 | 0.531 | 0.531 | 0.352 | 0.053 | 0.974 |
| 2 | 0.204 | 0.385 | 0.373 | 0.073 | 0.931 |
| 3 | 0.0957 | 0.469 | 0.492 | 0.215 | 0.768 |
| 4 | 0.0206 | 0.215 | 0.163 | 0.048 | 0.970 |
| 5 | $6.71 \times 10^{-3}$ | 0.325 | 0.328 | 0.102 | 0.909 |
| 6 | $1.20 \times 10^{-3}$ | 0.179 | 0.212 | 0.042 | 0.960 |
| 7 | $5.65 \times 10^{-4}$ | 0.470 | 0.383 | 0.147 | 0.869 |

The run used 351 jobs, 1.90 million shots and 1,440 s (24 min) of charged QPU time, against a predicted median of 1,442 s. We replayed the saved raw counts through the unchanged functions of the training and the refinement, and the results agreed bit for bit with the recorded run. We also recomputed the residual of every correction from the recorded pressure. With classical training, the same problem reaches $10^{-3}$ in four to five corrections at one job each (pilot of Section 4.1). The on-device training removes the classical simulation of the circuit, at a cost of 50 jobs per correction.

# 5. Discussion

## 5.1 Why the single solve stops

In a single solve, the direction error set by the state infidelity becomes the error of the pressure. Most of this error is a bias from gate and readout noise. The shot noise decreases as $1/\sqrt{N_s}$, but the bias does not. This is consistent with the fact that our single solves stopped at 0.33–0.59 despite 34,000 shots per circuit, and that Song et al. [2025] observed no convergence when the number of shots was raised from $10^5$ to $10^8$.

Two further features of the single-solve scheme of Song et al. add to this floor. First, the scale is set by the ratio of the largest elements of $A|\psi\rangle$ and $|b\rangle$, so that the noise in a single element sets the size of the whole solution, and the residual is not minimized. Second, the pressure null space is fixed by replacing a row of the operator, which changes the operator itself rather than only its null-space component. When the Poisson operator and the velocity correction do not use the same discrete divergence and gradient, the pressure residual and the physical divergence are not tied as in the identity of Section 2.1.

## 5.2 Why refinement converges

In the refinement scheme, the same quantum error enters as the factor $\sin\theta_k$ in each correction. A state fidelity of 0.97–0.98 is far below the $1 - F \leq 10^{-7}$ estimated by Song et al., yet it is sufficient for a contraction of about 0.1–0.2 per correction. At six qubits, a fidelity of about 0.965 still gave a contraction of about 0.25 per correction. The least-squares scale guarantees that the residual does not increase, and the MAC identity makes each reduction of the residual a reduction of the divergence by the same factor. This distinction separates the accuracy of one quantum solve from the accuracy of the final pressure.

The on-device run tests this argument in a stricter setting. With the training on hardware, the fidelity of the trained state fell to 0.77–0.97, and the contraction rose to about 0.32 per correction. The refinement still converged, at the rate predicted by the noise model. Each correction needs only a moderate accuracy, so that two Rotosolve sweeps with 500 shots per circuit are sufficient. In this sense, the refinement relaxes the accuracy required of each variational solve, and this makes on-device training practical. Song et al. [2025] did not train on the device because the hardware noise drifts with time and the training takes long. In the present run, the calibration of the qubit chain changed twice, and the refinement converged without retuning.

The formalism is general enough to apply to other symmetric positive linear systems in time-dependent PDE solvers, provided a classical residual and a variational state preparation for the normalized residual are available.

## 5.3 Scope and outlook

The present results define the following scope. (1) The results are for four and six qubits (16 and 64 unknowns) and a single time step with a fixed right-hand side, and the six-qubit result rests on three pairs. (2) In the main experiments, training uses an exact classical state-vector simulation, whose cost grows exponentially with the number of qubits; this limitation is shared with Song et al. The on-device training removes the classical simulation. It was demonstrated in one run, at four qubits and to a tolerance of $10^{-3}$, and its cost grows with the number of parameters, as two cost evaluations per angle and sweep. (3) HTree reconstructs all $2^n$ amplitudes, so that the number of shots needed grows with the problem size. (4) On IonQ, the present runs used no readout calibration and no matched-shot control; a matched-shot experiment on IonQ is in progress. (5) The aim of the present work is convergence on hardware; a comparison of the cost with classical solvers requires larger problems.

Two extensions build on these results. Time stepping with the energy-based tolerance of Section 2.5 is being tested over a 20-step startup trajectory. For larger problems, the quantum directions of successive corrections can be combined in a flexible Krylov subspace, and the readout can be replaced by measured projected quantities or by sample-based subspace selection.

# 6. Conclusions

In the present work, we combined a variational pressure solve on noisy quantum hardware with residual-based iterative refinement and a least-squares scale fitted to the physical residual. It is shown that (1) the error of each quantum solve enters as a contraction factor $\sin\theta_k$ rather than as an error floor; (2) with a boundary-consistent MAC discretization, each correction reduces the velocity divergence by the same factor as the residual; (3) on ibm_cleveland the relative residual of a four-qubit problem reached below $10^{-7}$ within eight to ten corrections in all 20 runs of a pre-registered experiment, whereas a single solve with a comparable total number of shots stopped at 0.33–0.59, and a six-qubit pilot reached the same tolerance in 13 corrections in all three runs; (4) on the IonQ Forte trapped-ion processor the same contraction behavior was observed, and a manufactured-solution benchmark reached the tolerance of $10^{-7}$ in nine corrections with a pressure error of $4.9 \times 10^{-8}$; and (5) with the variational training carried out on ibm_cleveland, the residual fell below $10^{-3}$ in seven corrections from zero initial parameters, at the rate predicted by a noise model. The results show that current NISQ hardware can support a converging pressure solve when the quantum step is used as an inexact solver inside a classical refinement. The method requires no change of the quantum circuit, and it carries over from four to six qubits and from superconducting to trapped-ion hardware.

# Acknowledgments

This work was supported by Korea National Research Foundation (NRF) grant No. NRF-2026-25612745, ICT R&D program of IITP-2023-2021-0-01810, RS-2023-00225385, AFOSR grant FA2386-22-1-4052, the Cleveland Clinic Quantum Innovation Catalyzer Program and Amazon Web Services. We acknowledge the use of IBM Quantum services for this work. The views expressed are those of the authors, and do not reflect the official policy or position of IBM or the IBM Quantum team. This work was also supported by the National Quantum Laboratory at Maryland (QLab). We acknowledge QLab for providing quantum computing resources on IonQ devices for this work. The views expressed are those of the authors, and do not reflect the official policy or position of IonQ. D. Ahn is a co-founder of Singularity Quantum Inc. and holds an equity interest in the company. The other authors declare no competing interests.

# Data availability

Raw IBM counts, IonQ probability artifacts, job identifiers, trained parameters and the analysis code will be deposited in Figshare upon publication. The repository also contains the bound native circuit of every job.

# Declaration of generative AI and AI-assisted technologies in the manuscript preparation process

During the preparation of this work, the authors used Claude (Anthropic) and ChatGPT (OpenAI) to assist in drafting and editing the text. After using these tools, the authors reviewed and edited the content as needed and take full responsibility for the content of the published article.

# Appendix A. Governing equations and derivations

This appendix collects the equations of the projection method in the form used in Section 2. The continuous equations and the fractional steps follow Song et al. [2025] and Chorin [1968]; the discrete identities are derived here for the boundary-consistent MAC discretization.

## A.1 Navier–Stokes equations and the projection method

The incompressible Navier–Stokes equations express the conservation of mass and momentum. For a fluid of constant density $\rho$ and kinematic viscosity $\nu$, they read

$$\nabla \cdot \mathbf{u} = 0, \tag{A.1}$$

$$\frac{\partial \mathbf{u}}{\partial t} + (\mathbf{u} \cdot \nabla)\mathbf{u} = -\frac{1}{\rho}\nabla p + \nu \nabla^2 \mathbf{u}, \tag{A.2}$$

where $\mathbf{u} = (u, v)$ is the velocity and $p$ is the pressure. In dimensionless form $\nu$ is replaced by 1/Re, where Re is the Reynolds number. We consider a closed domain $\Omega$ with a prescribed wall velocity $\mathbf{u} = \mathbf{u}_w$ on $\partial\Omega$ and $\mathbf{u}_w \cdot \mathbf{n} = 0$, as in the lid-driven cavity of Song et al. [2025].

The pressure has no evolution equation of its own. It acts as a Lagrange multiplier that enforces (A.1). By taking the divergence of (A.2) and using (A.1), we obtain

$$\nabla^2 p = -\rho \, \nabla \cdot [(\mathbf{u} \cdot \nabla)\mathbf{u}], \tag{A.3}$$

so that the pressure is a nonlocal functional of the instantaneous velocity [Doering and Gibbon 1995].

In the projection method of Chorin [1968], the time derivative is discretized by the explicit Euler method, and one step from $t^n$ to $t^{n+1} = t^n + \Delta t$ is split into three fractional steps. First, an intermediate velocity is computed without the pressure gradient,

$$\frac{\mathbf{u}^* - \mathbf{u}^n}{\Delta t} = -(\mathbf{u}^n \cdot \nabla)\mathbf{u}^n + \nu \nabla^2 \mathbf{u}^n. \tag{A.4}$$

The velocity $\mathbf{u}^*$ is not divergence free in general. Second, the pressure is obtained by requiring that the corrected velocity be divergence free,

$$\nabla^2 p^{n+1} = \frac{\rho}{\Delta t} \, \nabla \cdot \mathbf{u}^* \ \text{ in } \Omega, \qquad \frac{\partial p^{n+1}}{\partial n} = 0 \ \text{ on } \partial\Omega. \tag{A.5}$$

Third, the velocity is corrected by the pressure gradient,

$$\mathbf{u}^{n+1} = \mathbf{u}^* - \frac{\Delta t}{\rho} \, \nabla p^{n+1}. \tag{A.6}$$

The sum of (A.4) and (A.6) recovers the explicit discretization of (A.2), and the divergence of (A.6) together with (A.5) gives $\nabla \cdot \mathbf{u}^{n+1} = 0$. The Neumann condition in (A.5) follows from the normal component of (A.6) at the wall, where $\mathbf{u}^* \cdot \mathbf{n}$ and $\mathbf{u}^{n+1} \cdot \mathbf{n}$ both equal the prescribed value. The explicit time step is restricted by the advective (CFL) and viscous conditions

$$\Delta t \sum_k \frac{\| \mathbf{u} \|_{\max}}{\Delta x_k} \le 1, \qquad \nu \, \Delta t \sum_k \frac{1}{\Delta x_k^2} \le \frac{1}{2}, \tag{A.7}$$

where $\Delta x_k$ is the grid spacing in the $k$th direction. The first condition is the one used by Song et al. [2025]; the second is the standard stability condition for explicit diffusion.

## A.2 Staggered (MAC) discretization

On the MAC grid, the pressure $p_{i,j}$ is stored at the cell centers, the velocity component $u_{i+1/2,j}$ at the vertical faces and $v_{i,j+1/2}$ at the horizontal faces. The discrete divergence $D$ maps face velocities to cell centers, and the discrete gradient $G$ maps cell pressures to interior faces,

$$(Du)_{i,j} = \frac{u_{i+1/2,j} - u_{i-1/2,j}}{\Delta x} + \frac{v_{i,j+1/2} - v_{i,j-1/2}}{\Delta y}, \qquad (Gp)_{i+1/2,j} = \frac{p_{i+1,j} - p_{i,j}}{\Delta x}, \tag{A.8}$$

and similarly for the $y$ component of $Gp$. At a wall face the normal velocity is prescribed and is not corrected, so that $G$ is defined only on interior faces. On a uniform grid it can be shown that $G = -D^{\mathsf{T}}$ on the interior faces. Therefore $L = -DG = DD^{\mathsf{T}}$ is symmetric and positive semidefinite. For an interior cell,

$$(Lp)_{i,j} = \frac{2p_{i,j} - p_{i-1,j} - p_{i+1,j}}{\Delta x^2} + \frac{2p_{i,j} - p_{i,j-1} - p_{i,j+1}}{\Delta y^2}. \tag{A.9}$$

For a cell adjacent to a wall, the coefficient 2 is reduced to 1 and the term of the missing neighbor is omitted in that direction. This is the discrete form of the Neumann condition in (A.5). By applying $D$ to the discrete form of (A.6) and requiring $Du^{n+1} = 0$, we obtain the discrete Poisson equation $Lp = b$ with $b = -(\rho/\Delta t)Du^*$ of Section 2.1.

## A.3 Null space and the residual–divergence identity

Each row of (A.9) sums to zero, so that the constant vector $\mathbb{1}$ spans the null space of $L$. Since $L$ is symmetric, $Lp = b$ is solvable only if $\mathbb{1}^{\mathsf{T}}b = 0$. By the discrete divergence theorem, $\mathbb{1}^{\mathsf{T}}Du^*$ is proportional to the net volume flux through the walls, which vanishes for $\mathbf{u}_w \cdot \mathbf{n} = 0$.

We fix the additive constant with $A = L + e_0 e_0^{\mathsf{T}}/\Delta x^2$. Let $Ap = b$ with $\mathbb{1}^{\mathsf{T}}b = 0$. By multiplying by $\mathbb{1}^{\mathsf{T}}$ and using $\mathbb{1}^{\mathsf{T}}L = 0$, we obtain $p_0/\Delta x^2 = 0$. Therefore $p_0 = 0$ and $Lp = b$. That is, $A$ selects the solution of $Lp = b$ with $p_0 = 0$ and leaves the equation itself unchanged. In the same way, a direction $d^0$ with $d^0[0] = 0$ satisfies $Ad^0 = Ld^0$.

For any pressure $p$, not necessarily exact, the projected velocity $u = u^* - (\Delta t/\rho)Gp$ has the divergence

$$Du = Du^* - \frac{\Delta t}{\rho}DGp = -\frac{\Delta t}{\rho}\, b + \frac{\Delta t}{\rho}\, Lp = -\frac{\Delta t}{\rho}\,(b - Lp) = -\frac{\Delta t}{\rho}\, r. \tag{A.10}$$

This is the identity of Section 2.1. It holds for any $p$ and does not depend on the accuracy of the quantum solve. It requires only that the same $D$ and $G$ be used in the Poisson operator and in the correction step.

## A.4 Kinetic-energy balance and the spurious pressure work

Let $E = (\rho/2)\int_\Omega |\mathbf{u}|^2 \; d^d x$. For a divergence-free velocity with $\mathbf{u} = 0$ on $\partial\Omega$, the inner product of (A.2) with $\rho\mathbf{u}$ gives [Doering and Gibbon 1995]

$$\frac{dE}{dt} = -\rho\nu \;\| \nabla\mathbf{u} \|^2. \tag{A.11}$$

For a moving wall, the work of the viscous stress on the boundary is added to (A.11). The pressure term in (A.2) contributes $-\int \mathbf{u} \cdot \nabla p \, d^d x = \int p \, \nabla \cdot \mathbf{u} \, d^d x$, after integration by parts with $\mathbf{u} \cdot \mathbf{n} = 0$ on $\partial\Omega$. It vanishes for a divergence-free velocity, so that the pressure does no net work. When the pressure solve is inexact, the divergence (A.10) is nonzero, and the pressure term acts as a spurious power $P_s = \int p \, \nabla \cdot \mathbf{u} \, d^d x$. With (A.10) and the Cauchy–Schwarz inequality, we obtain

$$|P_s| \leq \frac{\Delta t}{\rho} \parallel p \parallel \parallel r \parallel, \tag{A.12}$$

where the discrete norms include the cell volume. Requiring $|P_s| \leq \varepsilon_E\, \rho\nu \parallel \nabla\mathbf{u} \parallel^2$ gives the tolerance of Section 2.5. The advective term also contributes $-(\rho/2)\int (\nabla\cdot\mathbf{u})|\mathbf{u}|^2\, d^d x$ for a non-solenoidal field. It is proportional to the same divergence, and we do not include it in the tolerance.

## A.5 Effective Hamiltonian and the contraction factor

For a nonsingular $A$ and $\hat{b} = b/\parallel b \parallel$, the effective Hamiltonian of Section 2.2 is [Liu et al. 2021; Song et al. 2025]

$$H = A^{\mathsf{T}}\big(I - |\hat{b}\rangle\langle\hat{b}|\big)A. \tag{A.13}$$

$H$ is positive semidefinite. Its ground state, with eigenvalue zero, is $|x\rangle \propto A^{-1}|b\rangle$, since $I - |\hat{b}\rangle\langle\hat{b}|$ annihilates $A|x\rangle \propto|\hat{b}\rangle$. For a normalized real trial state $\psi$, we obtain

$$\langle\psi|H|\psi\rangle = \parallel A\psi \parallel^2 - \big(\hat{b}\cdot A\psi\big)^2 = \parallel A\psi \parallel^2 \sin^2\varphi, \tag{A.14}$$

where $\varphi$ is the angle between $A\psi$ and $b$. The normalized energy $\langle\psi|H|\psi\rangle/\parallel A\psi \parallel^2$ is therefore $\sin^2\varphi$. In the refinement the right-hand side is the residual $r_k$. Since $Ad^0 = Ld^0$ (Appendix A.3), the contraction factor $\sin\theta_k$ of Section 2.4 is the same quantity, evaluated for the measured direction $d^0$ instead of the trained state. The training thus minimizes the contraction factor of the ideal state, and the contraction on hardware differs from it by the error of preparation and readout.

With the left preconditioner $M^{-1}$ of Song et al. [2025], (A.13) is formed with $\tilde{A} = M^{-1}A$ and $\tilde{b} = M^{-1}b$. The ground state is unchanged, and $\varphi$ is measured after applying $M^{-1}$. The preconditioner therefore changes the conditioning of the optimization but not the solution.

# Appendix B. Circuits executed on hardware

This appendix gives the circuits in the form in which they were executed. The logical circuit of Fig. 1 is the same in all experiments. Its entangling gates act on a linear chain, which matches a linear chain of physical qubits on both processors. The circuits were therefore compiled without SWAP gates.

On ibm_cleveland, each $R_y$ rotation compiles to $R_z$ and $\sqrt{X}$ (SX) gates, and each CNOT to a CZ gate with single-qubit rotations. The Hadamard gate of an HTree circuit is absorbed into the final single-qubit rotations, so that the five circuits of one correction have the same gate counts. On IonQ Forte-1, the circuits compile to the native GPI and GPI2 single-qubit gates and the ZZ two-qubit gate. Table B1 lists the gate counts of the submitted circuits.

Table B1. Gate counts of the circuits executed on hardware. The counts are those of the submitted native circuits of one correction; they are the same for all corrections of an experiment unless a range is given.

| Problem | Processor and qubits | Two-qubit gates | Single-qubit gates | Depth |
|---|---|---|---|---|
| P1, pre-registered and on-device runs (Secs. 3.2, 3.5) | ibm_cleveland, 21–24 and 9–12 | 15 CZ | 78 $R_z$, 58 SX | 56 |

| Problem | Processor and qubits | Two-qubit gates | Single-qubit gates | Depth |
|---|---|---|---|---|
| P3 (Sec. 3.4) | ibm_cleveland, 21–26 | 30 CZ | 144 $R_z$, 108 SX | 73 |
| P1 (Sec. 3.3) | IonQ Forte-1 | 15 ZZ | 336–342 GPI and GPI2 | — |
| P2 (Sec. 3.3) | IonQ Forte-1 | 15 ZZ | 56–58 GPI2 | — |

The two IonQ problems have the same number of ZZ gates. The trained angles of P2 compile to far fewer single-qubit gates.

Fig. B1 shows the Z-basis circuit submitted for the final readout of the first correction of the on-device run (job db8eilmil99s73fpr5j0, 4,000 shots). Its angles are the trained parameters of Table B2. The four circuits with a Hadamard gate differ only in the final rotations. The bound circuits of every job, in QPY format for IBM and in the native JSON format for IonQ, are included in the data repository.

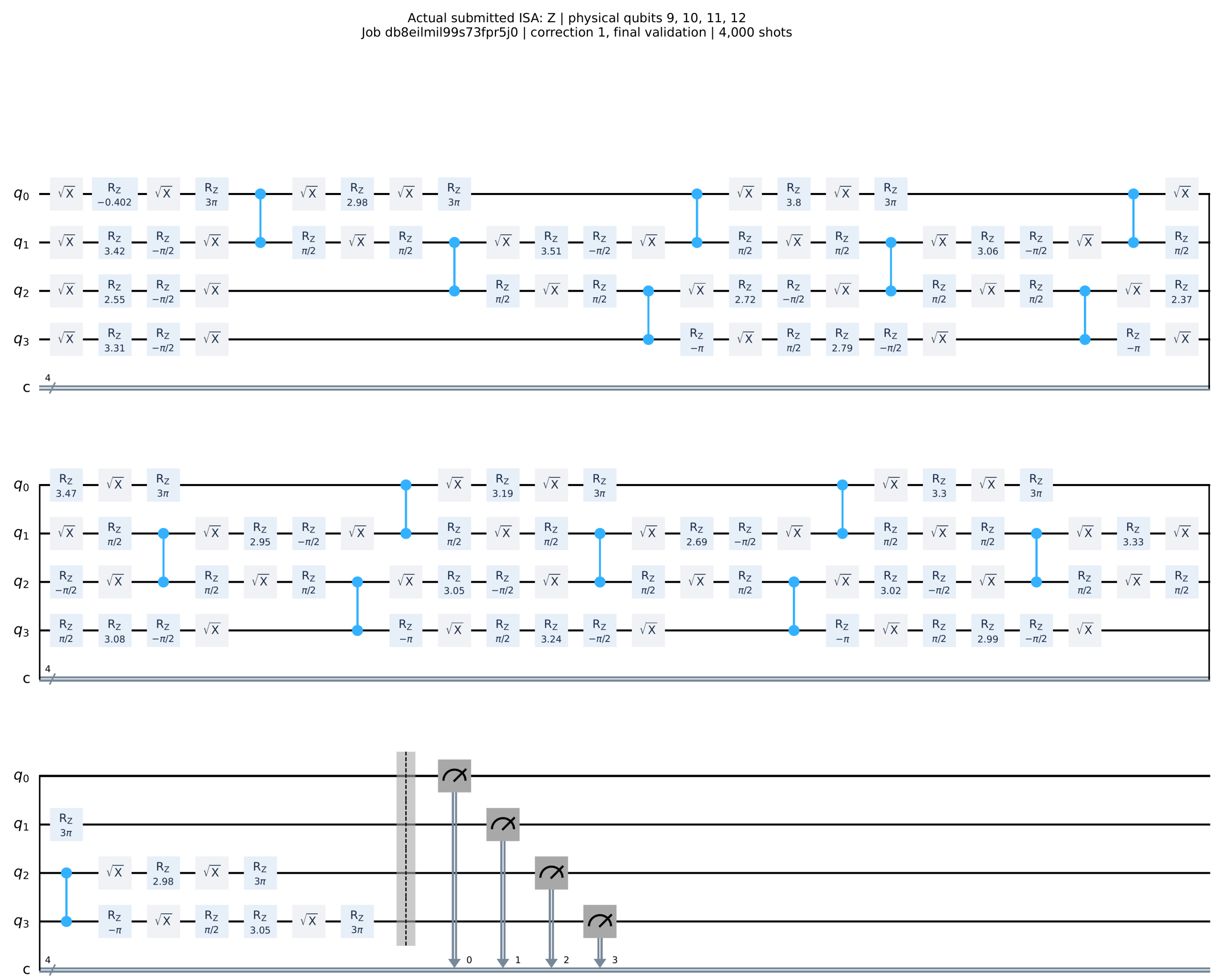

*Fig. B1. Native circuit submitted to ibm_cleveland for the Z-basis readout of the first correction of the on-device run (job db8eilmil99s73fpr5j0, 4,000 shots). The wires q0–q3 are the physical qubits 9–12. Idle physical qubits are omitted; the gate order, the bound angles (Table B2) and the measurements are as submitted. The circuit contains 15 CZ, 78 Rz and 58 SX gates, with a depth of 56.*

Table B2. Trained parameters $\theta_{4l+q}$ (rad) of the first on-device correction, bound to the circuit of Fig. B1. Here $l$ denotes the rotation layer and $q$ the qubit.

| Layer $l$ | $q = 0$ | $q = 1$ | $q = 2$ | $q = 3$ |
|---|---|---|---|---|
| 0 | −3.5433 | 0.2739 | −0.5901 | 0.1654 |
| 1 | −0.1597 | 0.3721 | −0.4224 | −0.3544 |
| 2 | 0.6617 | −0.0809 | −0.7757 | −0.0625 |
| 3 | 0.3299 | −0.1868 | −0.0925 | 0.0957 |
| 4 | 0.0500 | −0.4474 | −0.1170 | −0.1553 |
| 5 | 0.1585 | 0.1890 | −0.1577 | −0.0897 |

# References

1. A. J. Chorin, Numerical solution of the Navier–Stokes equations, Math. Comp. 22, 745–762 (1968).
2. C. R. Doering and J. D. Gibbon, Applied Analysis of the Navier–Stokes Equations (Cambridge University Press, 1995).
3. Z. Song, R. Deaton, B. Gard, and S. H. Bryngelson, Incompressible Navier–Stokes solve on noisy quantum hardware via a hybrid quantum–classical scheme, Comput. Fluids 288, 106507 (2025).
4. H.-L. Liu et al., Variational quantum algorithms for the Poisson equation, Phys. Rev. A 104, 022418 (2021).
5. C. Bravo-Prieto, R. LaRose, M. Cerezo, Y. Subasi, L. Cincio, and P. J. Coles, Variational quantum linear solver, Quantum 7, 1188 (2023).
6. J. H. Wilkinson, Rounding Errors in Algebraic Processes (Prentice-Hall, 1963).
7. E. Carson and N. J. Higham, Accelerating the solution of linear systems by iterative refinement in three precisions, SIAM J. Sci. Comput. 40, A817–A847 (2018).
8. E. Esmaeilifar, D. Ahn, and R. S. Myong, Quantum algorithm for nonlinear Burgers' equation for high-speed compressible flows, Phys. Fluids 36, 106110 (2024).
9. K. M. Nakanishi, K. Fujii, and S. Todo, Sequential minimal optimization for quantum-classical hybrid algorithms, Phys. Rev. Research 2, 043158 (2020).
10. M. Ostaszewski, E. Grant, and M. Benedetti, Structure optimization for parameterized quantum circuits, Quantum 5, 391 (2021).
11. M. Ali and M. Kabel, Performance study of variational quantum algorithms for solving the Poisson equation on a quantum computer, Phys. Rev. Applied 20, 014054 (2023).
12. G. Tosti Balducci, B. Chen, M. Möller, and R. De Breuker, Solving 1D Poisson problem with a variational quantum linear solver, arXiv:2412.04938 (2024).
13. Y. Saito, X. Lee, D. Cai, J. Shin, and N. Asai, Iterative refinement for variational quantum linear solver, in Proceedings of International Conference on Data Analytics and Insights, ICDAI 2023, Lecture Notes in Networks and Systems 727, 15–27 (Springer, 2023).
14. M. A. Schalkers and M. Möller, Efficient and fail-safe quantum algorithm for the transport equation, J. Comput. Phys. 502, 112816 (2024).
15. Z. Meng, L. Chen, J.-P. Liu, and G. He, Toward end-to-end quantum simulation of rapidly distorted turbulence, J. Comput. Phys. 558, 114888 (2026).
16. S. M. A. A. Mamun, S. Wang, J. J. Kath, D. O'Malley, Q. Kang, and Z. Li, Quantum unitary lattice Boltzmann formulation for canonical flow and thermal transport simulation, J. Comput. Phys. 568, 115381 (2027).
17. S. Jin, N. Liu, and Y. Yu, Time complexity analysis of quantum algorithms via linear representations for nonlinear ordinary and partial differential equations, J. Comput. Phys. 487, 112149 (2023).
18. S. Jin, X. Li, N. Liu, and Y. Yu, Quantum simulation for partial differential equations with physical boundary or interface conditions, J. Comput. Phys. 498, 112707 (2024).